\documentclass[aps,pra,twocolumn,showpacs]{revtex4}
\usepackage{graphicx}
\usepackage{psfrag}
\usepackage{amsmath}
\usepackage{amssymb}

\renewcommand{\vec}[1]{\boldsymbol{#1}}
\renewcommand{\tensor}[1]{{\boldsymbol{\mathit {#1}}}}
\newcommand{\Nabla}{\vec{\nabla}}
\newcommand{\Lnabla}{\overset{\leftarrow}{\Nabla}}

\newcommand{\EW}[1]{\langle #1 \rangle}

\renewcommand{\Im}{{\rm Im}\,}
\renewcommand{\Re}{{\rm Re}\,}
\newcommand{\uv}[1]{\mathbf{e}_{#1}}

\newcommand{\hank}{\tilde}
\newcommand{\ccangles}{\bullet}
\newcommand{\fo}[1]{\underline{\hat{\mathbf{#1}}}}
\begin{document}
\title{Casimir stress on lossy magnetodielectric spheres}
\author{Christian Raabe}
\author{Ludwig Kn\"{o}ll}
\author{Dirk-Gunnar Welsch}
\affiliation{Theoretisch-Physikalisches Institut,
Friedrich-Schiller-Universit\"at Jena, Max-Wien-Platz 1, D-07743 Jena,
Germany} 
\date{\today}
\begin{abstract}
An expression for the Casimir stress
on arbitrary dispersive and lossy linear
magnetodielectric matter at finite temperature,
including left-handed material, is
derived and applied to spherical systems.
To cast the relevant
part of the scattering Green tensor for a
general magnetodielectric sphere in a
convenient form, classical Mie scattering is reformulated.
\end{abstract}
\maketitle

\section{Introduction}
Spheres and spherical shells display enough
symmetry to allow for a
quite explicit solution of the Casimir problem
(for the earliest calculation, see \cite{Boyer}).
Especially two intriguing
but nowadays obsolete ideas have substantially
stimulated the study of 
spherically symmetric setups.
Firstly, the idea
that the Casimir effect might be responsible
for the postulated Poincar\'{e} stress
that miraculously holds the charge distribution of a classical
electron together \cite{CasimirElectronModel},
and secondly the idea that it
might be related \cite{SchwingerSonoluminescence} to sonoluminescence
\cite{SonoluminescenceFirst,SonoluminescenceReview}, i.e.,
the emission of light flashes by
small air bubbles in water under the action of ultrasonic
waves.

Unfortunately, in
the available work on the topic
(see, e.g.,
\cite{MiltonSonoluminescence,Barton,Klich1,Klich2,Brevik,Nesterenko2})
lossless materials are assumed, dispersion
is ignored, or the influence of matter is summarized
in perfect conductor boundary conditions
\cite{Boyer,MiltonDdim,Nesterenko}. Needless
to say that such approximations should be expected to 
distort the physical picture
more or less drastically,
not to mention their tendency
to cause `unreal' divergence problems.
In what follows, we allow for
dispersing and absorbing (causal) sphere material
characterized by both a complex frequency-dependent permittivity
and a complex frequency-dependent permeability, thus including  
in the calculations also 
left-handed materials, which have been
of increasing interest.
To our knowledge, Casimir forces on magnetodielectric bodies
have been studied only for non-dispersing and non-absorbing material  
\cite{MiltonSonoluminescence}, commonly subject
to the condition that the product of the permittivity and the
permeability is also uniform in
space \cite{Klich1,Klich2,Brevik,Nesterenko2}.
By using the quantization scheme given in 
Ref.~\cite{Welsch} in our calculations,
we first extend the
recently derived basic formula \cite{Raabe1} for the Casimir
force for causal dielectric matter to causal magnetodielectric matter.
We then apply the theory to spherical structures.

\section{Quantization scheme}
Within the mentioned
quantization scheme, the macroscopic (medium-assisted)
electromagnetic field operators are all expressed in
terms of suitable bosonic basic fields via
the classical Green tensor, which satisfies the equation
\begin{equation}
\label{2.1}
\Nabla\times\kappa(\mathbf{r},\omega)\Nabla\times
\tensor{G}(\mathbf{r,r'},\omega)
- \frac{\omega^2}{c^2}\varepsilon(\mathbf{r},\omega)
\tensor{G}(\mathbf{r,r'},\omega)
\!=\!\tensor{\delta}(\mathbf{r,r'})
\end{equation}
($\varepsilon$ -- complex permittivity, $\kappa=\mu^{-1}$ --
complex reciprocal permeability) and the boundary condition at infinity.
In the absence of additional charges and currents, the
positive-frequency components of the electric field operator
are then given by
\begin{equation}
\label{2.2}
\fo{E}(\mathbf{r},\omega)=i\mu_{0}\omega\int
d^3r'\,\tensor{G}(\mathbf{r,r'},\omega)
\fo{j}_{\mathrm N}(\mathbf{r},\omega),
\end{equation}
from which the other field operators
can be derived by using Faraday's law
\begin{equation}
\label{2.2a}
\fo{B}(\mathbf{r,\omega})=
(i\omega)^{-1}\Nabla\times\fo{E}(\mathbf{r},\omega)
\end{equation}
and the constitutive relations [$\kappa_{0}\!=\!\mu_{0}^{-1}$]
\begin{eqnarray}
\label{2.3}
\fo{D}(\mathbf{r},\omega)&=&\varepsilon_{0}
\varepsilon(\mathbf{r},\omega)\fo{E}(\mathbf{r},\omega)
+\fo{P}_\mathrm{N}(\mathbf{r},\omega),\\\nonumber
\fo{H}(\mathbf{r},\omega)&=&\kappa_{0}\kappa(\mathbf{r},\omega)\fo{B}
(\mathbf{r},\omega)-\fo{M}_\mathrm{N}(\mathbf{r},\omega).
\end{eqnarray}
In the above, the frequency-domain Langevin noise quantities
(carrying an index N) are connected with the fundamental
bosonic fields $\hat{\mathbf{f}}_{\lambda}(\mathbf{r},\omega)$
($\lambda\!=\!e,m$) by the relations
\begin{eqnarray}
\label{2.4}
\fo{j}_\mathrm{N}(\mathbf{r},\omega)&=&-i\omega
\fo{P}_\mathrm{N}(\mathbf{r},\omega)
+\Nabla\times\fo{M}_\mathrm{N}(\mathbf{r},\omega),
\\
\label{2.4-1}
\fo{P}_\mathrm{N}(\mathbf{r},\omega)
&=&i\left[\hbar\varepsilon_{0}
\Im\varepsilon(\mathbf{r,\omega})/\pi\right]^{1/2}
\hat{\mathbf{f}}_{e}(\mathbf{r},\omega),
\\
\label{2.4-2}
\fo{M}_\mathrm{N}(\mathbf{r},\omega)
&=&\left[-\hbar\kappa_{0}
\Im\kappa(\mathbf{r,\omega})/\pi\right]^{1/2}
\hat{\mathbf{f}}_{m}(\mathbf{r},\omega).
\end{eqnarray}
As a consequence, the
electromagnetic field operators satisfy the usual
commutation relations of QED.
For example, the electric field operator (in the
Schr\"{o}dinger picture) reads as
\begin{equation}
\label{2.4b}
\hat{\mathbf{E}}(\mathbf{r})
=\int_{0}^{\infty}d\omega\,\fo{E}(\mathbf{r,\omega})+ \mathrm{H.c.},
\end{equation}
with $\fo{E}(\mathbf{r,\omega})$ from Eq.~(\ref{2.2}) together
with Eqs.~(\ref{2.4})--(\ref{2.4-2}).
Writing down the corresponding formulas for the other field operators is
straightforward. The consistency of the method relies
heavily on the facts that the Green tensor, which is a response function
like $\varepsilon(\mathbf{r},\omega)$
and $\kappa(\mathbf{r},\omega)$, is holomorphic in the upper
$\omega$ half-plane, obeys the reciprocity relation 
\begin{equation}
\label{2.4c}
\tensor{G}(\mathbf{r,r'},\omega)
=\tensor{G}^\mathrm{T}(\mathbf{r',r},\omega)
\end{equation}
(the superscript T denotes matrix transposition), and the
integral relation
\begin{eqnarray}
\label{2.5}
\lefteqn{
\int d^3s\left\{
\bigl[\tensor{G}(\mathbf{r,s},\omega)\!\times\! \Lnabla_{\!s}\bigr]
\Im\kappa(\mathbf{s},\omega)
\bigl[\Nabla_{\!s}\!\times\!\tensor{G}^{\ast}(\mathbf{s,r'},\omega)\bigr]
\right.
}
\nonumber\\&&\hspace{-1ex}
\left.
+\,\frac{\omega^2}{c^2}\tensor{G}(\mathbf{r,s},\omega)
\Im\varepsilon(\mathbf{s,\omega})
\tensor{G}^{\ast}(\mathbf{s,r'},\omega)
\right\}\!=\!\Im\tensor{G}(\mathbf{r,r'},\omega)
\nonumber\\&&
\end{eqnarray}
(for details, see \cite{Welsch}). Note that the
convention
\begin{equation}
\label{2.6}
\tensor{G}(\mathbf{r,r'},\omega)\times\Lnabla{'}
=-\Nabla'\times \tensor{G}^\mathrm{T}(\mathbf{r,r'},\omega)
\end{equation}
has been used.

\section{Casimir stress tensor}
For a given quantum state, the stress tensor,
i.e., the (yet unrenormalized) Casimir stress,
can be obtained from
the correlation function
\begin{multline}
\label{2.7}
\tensor{T}(\mathbf{r,r'})
=\EW{\hat{\mathbf{D}}(\mathbf{r})\otimes\hat{\mathbf{E}}(\mathbf{r'})}
+\EW{\hat{\mathbf{B}}(\mathbf{r})\otimes\hat{\mathbf{H}}(\mathbf{r'})}
\\
-{\textstyle\frac{1}{2}}\,\tensor{1} {\rm
Tr}\!\left[\EW{\hat{\mathbf{D}}(\mathbf{r})
\otimes\hat{\mathbf{E}}(\mathbf{r'})}
+\EW{\hat{\mathbf{B}}(\mathbf{r})\otimes\hat{\mathbf{H}}(\mathbf{r'})}\right]
\end{multline}
in the coincidence
limit \mbox{$\mathbf{r'}$ $\!\to$ $\!\mathbf{r}$}.
In carrying out this limit, we have
to drop -- in accordance with the Casimir
effect's very definition -- the bulk part of the Green
tensor, which is singular but
independent of geometry \footnote{Identification of the bulk part is
unambiguously possible at any point in space where the material
properties are homogeneous, in particular in free-space regions.},
to get a physical (finite) value of the Casimir force per unit area.
Note that the resulting force formula rigorously applies only to
points outside matter, which is the case under consideration.

Let us calculate the thermal-equilibrium
Casimir force at temperature $T$. Recalling
the bosonic character of the fundamental fields
$\hat{\mathbf{f}}_{\lambda}(\mathbf{r,\omega})$
and assuming them to be excited in thermal states,
one quickly finds,
in close analogy to Ref.~\cite{Raabe1},
that
\begin{multline}
\label{2.8}
\bigl\langle\hat{\mathbf{f}}_{\lambda}(\mathbf{r,\omega})\otimes
\hat{\mathbf{f}}_{\lambda'}^{\dagger}
(\mathbf{r',\omega'})\bigr\rangle
\\
={\textstyle\frac{1}{2}}
\left[\coth\left(\frac{\hbar\omega}{2k_{B}T}\right)+1
\right]\delta_{\lambda\lambda'}\delta(\omega-\omega')
\tensor{\delta}(\mathbf{r,r'}),
\end{multline}
\begin{multline}
\label{2.9}
\bigl\langle\hat{\mathbf{f}}_{\lambda}^{\dagger}(\mathbf{r,\omega})\otimes
\hat{\mathbf{f}}_{\lambda'}
(\mathbf{r',\omega'})\bigr\rangle
\\
={\textstyle\frac{1}{2}}
\left[\coth\left(\frac{\hbar\omega}{2k_{B}T}\right)-1
\right]\delta_{\lambda\lambda'}\delta(\omega-\omega')
\tensor{\delta}(\mathbf{r,r'}),
\end{multline}
\begin{equation}
\label{2.10}
\bigl\langle\hat{\mathbf{f}}_{\lambda}(\mathbf{r,\omega})\otimes
\hat{\mathbf{f}}_{\lambda'}
(\mathbf{r',\omega'})\bigr\rangle
=\bigl\langle\hat{\mathbf{f}}_{\lambda}^{\dagger}
(\mathbf{r,\omega})
\otimes\hat{\mathbf{f}}_{\lambda'}
(\mathbf{r',\omega'})\bigr\rangle=0.
\end{equation}
By using Eq.~(\ref{2.4b})
for the electric field and the
related equations for the
other fields [together with Eqs.~(\ref{2.2})--(\ref{2.4-2})] and
Eqs.~(\ref{2.8})--(\ref{2.10}), and by employing the Green tensor
properties (\ref{2.4c}) and (\ref{2.5}),
it follows by a cumbersome but straightforward
calculation that Eq.~(\ref{2.7}) may be rewritten as
\begin{equation}
\label{2.11}
\tensor{T}(\mathbf{r,r'})=\tensor{\Theta}(\mathbf{r,r'})
-{\textstyle\frac{1}{2}}\tensor{1}{\rm Tr}\,\tensor{\Theta}(\mathbf{r,r'}),
\end{equation}
where
\begin{eqnarray}
\label{2.12}
\lefteqn{
\tensor{\Theta}(\mathbf{r,r'})=\frac{\hbar}{\pi}\int_{0}^{\infty}d\omega\,
\coth\!\left(\frac{\hbar\omega}{2k_{B}T}\right)
}
\nonumber\\&&\hspace{5ex}\times\,
\Im\!\biggl\{\frac{\omega^2}{c^2}\varepsilon(\mathbf{r},\omega)\tensor{G}
(\mathbf{r,r'},\omega)
\nonumber\\&&\hspace{9ex}
-\,\Bigl[\Nabla\!\times\!\tensor{G}(\mathbf{r,r'},\omega)
\times\Lnabla{'}\Bigr]\kappa(\mathbf{r'},\omega)
\biggr\}.
\end{eqnarray}
Note that this result differs
-- apart from the different analytical form of the Green tensor --
from the `non-magnetic' result \cite{Raabe1} in the
permeability that appears in the second term in the curly brackets. 
By replacing the full Green tensor with its scattering part,
\mbox{$\tensor{G}(\mathbf{r,r'},\omega)$ $\!\mapsto$
$\!\tensor{G}^\mathrm{scat}(\mathbf{r,r'},\omega)$},
and taking the coincidence limit \mbox{$\mathbf{r}'$ $\!\to$
$\!\mathbf{r}$}, Eq.~(\ref{2.11}) together with Eq.~(\ref{2.12})
is the sought-after basic formula for the Casimir force per unit
area for an arbitrary arrangement of lossy magnetodielectric bodies.
With the bulk Green tensor removed and application of
the standard rule $\omega\!\mapsto\!\omega\!+\!i0$ to handle
free-space regions, no divergence problems arise in performing the
integral in Eq.~(\ref{2.12}).

\section{Sphere Green tensor}
To apply the theory to spheres,
we need a convenient expression for the sphere scattering
Green tensor -- a purely classical problem,
which was first treated by Mie \cite{Mie} and
is now standard textbook material (see, e.g.,
\cite{ChewBook,BornWolf}). Here,
we give a formulation that is valid for arbitrary
complex permittivities and permeabilities that vary radially in
a stepwise fashion. (Note that amplifying materials are in fact also
allowed.) The required techniques can
be gathered from the mentioned volumes, but see also
\cite{Jackson,Taibook,MorseAndFeshbach}.
According to the method of Debye's potentials, transverse
solutions of the vector wave equation
\begin{equation}
\label{3.1}
\Nabla\times\Nabla\times\mathbf{F}(\mathbf{r})
-k^2\mathbf{F}(\mathbf{r})=0
\end{equation}
are given,
for each Debye potential $u$
that solves the scalar Helmholtz equation
\mbox{$(\Delta$ $\!+$ $\!k^2)u$ $\!=$ $\!0$},
by vector functions
\mbox{$\mathbf{F}$ $\!=$ $\!\Nabla\times\mathbf{r}u$} and 
\mbox{$\mathbf{F}$ $\!=$ $\!\Nabla\times\Nabla\times\mathbf{r}u$}.
In addition, all longitudinal vector functions
\mbox{$\mathbf{F}$ $\!=$ $\!\Nabla u$}
solve Eq.~(\ref{3.1}) for \mbox{$k$ $\!=$ $\!0$}. Note that
the free-space eigenfunctions of the (semi-bounded) operator
$\Nabla\!\times\!\Nabla\times$ obtained from
Eq.~(\ref{3.1}) under the restriction that they be bounded
everywhere (including infinity)
have real $k$ and are complete
\cite{MorseAndFeshbach}. In free space,
an orthonormalized and complete set of scalar (continuum)
eigenfunctions $u$ reads
\begin{equation}
\label{3.2}
u_{lm}(\mathbf{r}; k)
= \sqrt{\frac{2}{\pi}}\,k \,j_{l}(kr)Y_{lm}(\theta,\phi)
\quad
(k \ge 0), 
\end{equation}
with $l$ $\!=$ $\!0,1,2\ldots$, $m$ $\!=$ $\!-l,\ldots,l$
($j_{l}$ -- spherical Bessel function, $Y_{lm}$ -- spherical
harmonic \footnote{We employ standard definitions of these special
functions. For definiteness, we refer to Ref.~\cite{Jackson}.}).
Employing them as Debye potentials
and introducing appropriate normalization
factors yields the orthonormalized vector eigenfunctions
as
\begin{eqnarray}
\label{3.3}
\lefteqn{
\mathbf{L}_{lm}(\mathbf{r};k)=\frac{1}{k}\Nabla
u_{lm}(\mathbf{r};k)
}
\nonumber\\&&
\!=\sqrt{\frac{2}{\pi}}
\left[k j'_{l}(kr)\, \mathbf{r}Y_{lm}(\theta,\phi)/r
+ j_{l}(kr)\,\Nabla Y_{lm}(\theta,\phi)\right],\qquad
\end{eqnarray}
\begin{eqnarray}
\label{3.4}
\lefteqn{
\mathbf{M}_{lm}(\mathbf{r};k) = [l(l+1)]^{-\frac{1}{2}}\,
\Nabla\times\mathbf{r}u_{lm}(\mathbf{r};k)
}
\nonumber\\&&
= - \sqrt{\frac{2}{\pi l(l+1)}}\,k j_{l}(kr)\mathbf{r}\times
\Nabla Y_{lm}(\theta,\phi),\qquad
\end{eqnarray}
\begin{eqnarray}
\label{3.5}
\lefteqn{
\mathbf{N}_{lm}(\mathbf{r};k)
=\frac{1}{k}\Nabla\times\mathbf{M}_{lm}(\mathbf{r};k)
}
\nonumber\\&&
=\sqrt{\frac{2}{\pi l(l+1)}}\,
\left\{l(l+1)  j_{l}(kr)\,
\mathbf{r}Y_{lm}(\theta,\phi)/r^2
\right.
\nonumber\\&&\hspace{6ex}
\left.
+\left[k r j'_{l}(kr)+ j_{l}(kr)
\right] \Nabla Y_{lm}(\theta,\phi)\right\},
\end{eqnarray}
where for the $\mathbf{M}$ and $\mathbf{N}$ functions
the value $l$ $\!=$ $\!0$ has to be excluded from consideration.
Throughout the literature it is invariably assumed
that the Debye potential method does not miss any eigenfunctions.
We also do so and assume the completeness relations
\begin{equation}
\label{3.6}
\sum_{lm}\int_{0}^\infty dk\, \mathbf{L}_{lm}(\mathbf{r};k)\otimes
\mathbf{L}_{lm}^{\ast}(\mathbf{r'};k)
=\tensor{\delta}^{\parallel}(\mathbf{r,r'}),
\end{equation}    
\begin{multline}
\label{3.7}
\sideset{}{'}\sum_{lm}\int_{0}^\infty dk\,
\left[\mathbf{M}_{lm}(\mathbf{r};k)\otimes
\mathbf{M}_{lm}^{\ast}(\mathbf{r'};k)\right.\\\left.
+ \mathbf{N}_{lm}(\mathbf{r};k)\otimes\mathbf{N}_{lm}^{\ast}
(\mathbf{r'};k)\right]
=\tensor{\delta}^{\perp}(\mathbf{r,r'}),
\end{multline} 
where the primed sum begins with
\mbox{$l$ $\!=$ $\!1$}. Note that the $\mathbf{M}$ and $\mathbf{N}$
functions represent, respectively, 
TE and TM (to $\mathbf{r}$) partial waves,
and $\mathbf{r}u$ plays the role of a Hertz vector.

To find the sphere scattering Green tensor, we will
make use of the vector functions just constructed. For this purpose,
let us first
consider wave propagation in a homogeneous, isotropic bulk
material, so that Eq.~(\ref{2.1}) simplifies to
\begin{eqnarray}
\label{3.9}
\lefteqn{
\Nabla\times\Nabla\times\tensor{G}^{\rm (bulk)}
(\mathbf{r,r'},\omega)
}
\nonumber\\&&
-\,k^2(\omega)\tensor{G}^{\rm (bulk)}(\mathbf{r,r'},\omega)
=\mu(\omega)\tensor{\delta}(\mathbf{r,r'}),
\end{eqnarray}
with arbitrary complex
\begin{equation}
\label{3.8}
k^2(\omega)=\frac{\omega^2}{c^2}\,\varepsilon(\omega)\mu(\omega).
\end{equation}
The bulk-material Green tensor can then be expressed
straightforwardly in terms of the
vector functions (\ref{3.4}) and (\ref{3.5}) according to
\begin{eqnarray}
\label{3.10}
\lefteqn{
\tensor{G}^{\rm (bulk)}(\mathbf{r,r'},\omega)\mu^{-1}(\omega)
=-\frac{\tensor{\delta}^{\parallel}(\mathbf{r,r'})}{k^2(\omega)}
}
\nonumber\\&&
+\sideset{}{'}\sum_{lm}\int_{0}^\infty\!\! \frac{dk'}{k'^2-k^2(\omega)}\,
\left[\mathbf{M}_{lm}(\mathbf{r};k')\otimes
\mathbf{M}_{lm}^{\ast}(\mathbf{r'};k')
\right.
\nonumber\\&&\hspace{15ex}
\left.
+\mathbf{N}_{lm}(\mathbf{r};k')
\otimes\mathbf{N}_{lm}^{\ast}(\mathbf{r'};k')\right],
\end{eqnarray}
and the
integral can be evaluated by means of the residue theorem 
(with due care of `static'
poles in the $\mathbf{N}\otimes\mathbf{N}^\ast$ terms \cite{ChewBook})
to obtain \mbox{[$k$ $\!=$ $\!k(\omega)$]}
\begin{widetext}
\begin{multline}
\label{3.11}
\tensor{G}^{\rm (bulk)}(\mathbf{r,r'},\omega)\mu^{-1}(\omega)
=-\frac{\uv{r}\otimes\uv{r}}{k^2}\,\delta(\mathbf{r,r'})
+\frac{i\pi}{2k}\sideset{}{'}\sum_{lm}
\left\{
\left[\mathbf{M}_{lm}(\mathbf{r};k)\otimes
\mathbf{\hank M}_{lm}^{\ccangles}(\mathbf{r'};k)
+ \mathbf{N}_{lm}(\mathbf{r};k)
\otimes\mathbf{\hank N}_{lm}^{\ccangles}(\mathbf{r'};k)
\right]\,\theta(r'-r)
\right.
\\
\left.
+\left[\mathbf{\hank M}_{lm}(\mathbf{r};k)\otimes
\mathbf{M}_{lm}^{\ccangles}(\mathbf{r'};k)
+ \mathbf{\hank N}_{lm}(\mathbf{r};k)
\otimes\mathbf{N}_{lm}^{\ccangles}(\mathbf{r'};k)
\right]\,\theta(r-r')
\right\}.
\end{multline}
\end{widetext}
Here, the `bullet' symbol ($^{\scriptstyle\ccangles}$) denotes
complex conjugation of the $k$-independent factors only,
and the tilde symbol ($\,\hank{}\,$) means that
in the definitions (\ref{3.4}) and (\ref{3.5}) the
spherical Bessel functions $j_{l}$ have
to be replaced with the `outgoing' and `incoming' spherical Hankel functions
\mbox{$\hank h_{l}$ $\!=$ $\!h_{l}^{(1)}$} and
\mbox{$\hank h_{l}$ $\!=$ $\!h_{l}^{(2)}$} for 
\mbox{$\Im k$ $\!>$ $\!0$}
and \mbox{$\Im k$ $\!<$ $\!0$}, respectively, to ensure
amplitude decay as \mbox{$|\mathbf{r}$ $\!-$ $\!\mathbf{r'}|$
$\!\to$ $\!\infty$}. The advantage of this notation is that the sign
in \mbox{$k$ $\!=$ $\!\pm\omega\sqrt{\varepsilon\mu}/c$}
can be chosen freely. In this way, the formulas obtained for the
Green tensor apply also to left-handed
and amplifying material.

The terms within the curly brackets in Eq.~(\ref{3.11})
correspond to TE and TM partial waves that
are solutions to the homogeneous version of the
differential equation (\ref{3.9}) with the 
dispersion relation (\ref{3.8}).
In a spherically layered medium (in contrast to the bulk material
considered so far), additional
waves that arise from
reflection at and transmission through layer interfaces
must be taken into account. They just form
the scattering part of the Green tensor
that we are interested in. The total
field must satisfy the
well-known continuity conditions
at the layer interfaces, the required amplitude matching can be done
for each partial wave separately. In fact, only the solutions
\mbox{$\sim$ $\!k \mathcal{S}_{l}(kr)$}
of the radial Helmholtz equation $[\frac{\partial^2}{\partial r^2}$$
+\frac{2}{r}\frac{\partial}{\partial
r}$$-\frac{l(l+1)}{r^2}$$+k^2]$$\mathcal{S}_{l}(kr)$$=0$
are involved in this matching process. To be more specific, 
$\mathcal{S}_{l}(kr)$ must be
a (bounded) superposition of the different
kinds of spherical Bessel functions
(for complex $k$)
in every layer such that
\begin{itemize}
\item for TE waves continuity of 
\begin{equation}
\label{3.12}
kr\,\mathcal{S}_{l}(k
r) \quad\mbox{and}\quad \mu^{-1}k \frac{\partial}{\partial
r}\left[r \mathcal{S}_{l}(kr)\right]
\end{equation}
\item and for TM waves continuity of 
\begin{equation}
\label{3.13}
\varepsilon\, k r\, \mathcal{S}_{l}(kr) \quad\mbox{and}\quad
k \frac{\partial}{\partial
r}\left[r \mathcal{S}_{l}(kr)\right]
\end{equation}
\end{itemize}
is ensured at spherical interfaces
\mbox{$r$ $\!=$ $\!{\rm const}$}.

The simplest case is a
homogeneous sphere (index $1$) of radius $R$
in a homogeneous environment (index $2$).
If both $\mathbf{r}$ and $\mathbf{r}'$
are outside the
sphere and \mbox{$R$ $\!<$ $r$ $\!<$ $\!r'$}
is valid, it can be concluded
from the above mentioned conditions of continuity
that the reflection coefficients related to the terms with
\mbox{$\theta(r'$ $\!-$ $\!r)$} in Eq.~(\ref{3.11}) are given by   
\begin{equation}
\label{3.14}
r_{l,21}^{\rm TE}=-\frac{\frac{j_{l}(k_{1}R)}{\mu_{2}}
\frac{\partial [R j_{l}(k_{2}R)]}{\partial R}-
\frac{j_{l}(k_{2}R)}{\mu_{1}}
\frac{\partial [R j_{l}(k_{1}R)]}{\partial R}}
{\frac{j_{l}(k_{1}R)}{\mu_{2}}
\frac{\partial [R \hank h_{l}(k_{2}R)]}{\partial R}-
\frac{\hank h_{l}(k_{2}R)}{\mu_{1}}
\frac{\partial [R j_{l}(k_{1}R)]}{\partial R}}
\end{equation}
and
\begin{equation}
\label{3.15}
r_{l,21}^{\rm TM}=-\frac{\frac{j_{l}(k_{1}R)}{\varepsilon_{2}}
\frac{\partial [R j_{l}(k_{2}R)]}{\partial R}-
\frac{j_{l}(k_{2}R)}{\varepsilon_{1}}
\frac{\partial [R j_{l}(k_{1}R)]}{\partial R}}
{\frac{j_{l}(k_{1}R)}{\varepsilon_{2}}
\frac{\partial [R \hank h_{l}(k_{2}R)]}{\partial R}-
\frac{\hank h_{l}(k_{2}R)}{\varepsilon_{1}}
\frac{\partial [R j_{l}(k_{1}R)]}{\partial R}}
\end{equation}
for TE and TM waves, respectively. Introducing the definition
\begin{multline}
\label{3.16}
\mathcal{D}[j_{l}(k_{1}R),\hank h_{l}(k_{2}R);\alpha]
\\
=\det\!\left[
\begin{array}{cc}
j_{l}(k_{1}R) & \hank h_{l}(k_{2}R) \\
\alpha_{1}^{-1}\partial j_{l}(k_{1}R)/\partial R &\alpha_{2}^{-1}\partial
\hank h_{l}(k_{2}R)/\partial R
\end{array}\right],
\end{multline}
we may write Eqs.~(\ref{3.14}) and (\ref{3.15}) in the
compact form of
\begin{equation}
\label{3.17}
r_{l,21}^{\rm TE}=
-\frac{\mathcal{D}[Rj_{l}(k_{1}R),Rj_{l}(k_{2}R);\mu]
}{\mathcal{D}[Rj_{l}(k_{1}R),R\hank h_{l}(k_{2}R);\mu]}
\end{equation}
and
\begin{equation}
\label{3.18}
r_{l,21}^{\rm TM}=
-\frac{\mathcal{D}[Rj_{l}(k_{1}R),Rj_{l}(k_{2}R);\varepsilon]
}{\mathcal{D}[Rj_{l}(k_{1}R),R\hank h_{l}(k_{2}R);\varepsilon]}\,,
\end{equation}
respectively. Note that the so-called Mie resonances of
the sphere can be found by
studying these expressions in the complex $\omega$ plane. The
contributions to the scattering Green tensor
for the case \mbox{$r$ $\!>$ $\!r'$ $\!>$ $\!R$} [related to
the terms with \mbox{$\theta(r$ $\!-$ $\!r')$} in Eq.~(\ref{3.11})]
need not be figured out separately, but may also
be found from 
the condition that the scattering Green tensor has to
be continuously differentiable
at \mbox{$r$ $\!=$ $\!r'$} (since the bulk
part alone accounts for the necessary singularity).
The scattering Green tensor can therefore be given, for the
case that both arguments
$\mathbf{r}$ and $\mathbf{r}'$ are outside the sphere, in the form of
\begin{multline}
\label{3.19}
\!\!\tensor{G}^{(22)}(\mathbf{r,r'},\omega)
\!=\!\!\frac{i\pi \mu_{2}}{2k_{2}}\!\sideset{}{'}\sum_{lm}
\!\bigl\{
r_{l,21}^{\rm TE}\mathbf{\hank M}_{lm}(\mathbf{r};k_{2})\otimes
\mathbf{\hank M}_{lm}^{\ccangles}(\mathbf{r'};k_{2})
\\
+r_{l,21}^{\rm TM}\mathbf{\hank N}_{lm}(\mathbf{r};k_{2})\otimes
\mathbf{\hank N}_{lm}^{\ccangles}(\mathbf{r'};k_{2})
\bigr\}.
\end{multline}
By construction, $\tensor{G}^{(22)}(\mathbf{r,r'},\omega)$ is
continuously differentiable
at \mbox{$\mathbf{r}$ $\!=$ $\!\mathbf{r'}$}, vanishes
for \mbox{$|\mathbf{r}|$ $\!\to$ $\!\infty$} and/or
\mbox{$|\mathbf{r}'|$ $\!\to$ $\!\infty$},
and satisfies the condition of
reciprocity. Note that in fact
\mbox{$\tensor{G}^{(22)\ccangles}$ $\!=$ $\!\tensor{G}^{(22)}$},
because of the addition theorem
\begin{equation}
\label{3.20}
\sum_{m=-l}^{l}Y_{lm}(\theta,\phi)Y_{lm}^{\ast}(\theta',\phi')
=\frac{2l+1}{4\pi}\,P_{l}(\cos\psi) \,\, \in \mathbb{R}
\end{equation}
of the spherical harmonics ($\psi$ is the angle between
the primed and unprimed directions).
It is worth noting that the form of Eq.~(\ref{3.19}) does
not change when the sphere is not homogeneous but 
consists of radial layers. Clearly, the
reflection coefficients (\ref{3.17}) and (\ref{3.18})
must then be replaced by generalized ones
that take `subsurface' reflections into account. These can be
calculated recursively
by standard methods \cite{ChewBook}, on
the basis of the matching conditions (\ref{3.12}) and (\ref{3.13}).

\section{Casimir stress on a sphere}
With Eq.~(\ref{3.19}) at hand, we can now
evaluate
the stress tensor according to
Eqs.~(\ref{2.11}) and (\ref{2.12}) [we set
\mbox{$\tensor{G}$ $\!\mapsto$
$\!\tensor{G}^\mathrm{scat}$ $\!\mapsto$
$\!\tensor{G}^{(22)}$}, to approach the sphere surface
from outside], on using
the symmetry between the $\mathbf{M}$ and
$\mathbf{N}$ functions and assuming free space around the
sphere, i.e., $\varepsilon_{2}\!=\!\mu_{2}\!=\!1$,
$k_{2}\!=\!\omega/c$ $\!+$ $\!i0$:
\begin{equation}
\label{4.0}
\tensor{T}(\mathbf{r,r})=\tensor{\Theta}(\mathbf{r,r})
-{\textstyle\frac{1}{2}}\tensor{1}{\rm Tr}\,\tensor{\Theta}(\mathbf{r,r}),
\end{equation}
\begin{multline}
\label{4.1}
\tensor{\Theta}(\mathbf{r,r})\!=\!
\frac{\hbar}{2c} \int_{0}^\infty\!\!\!
d\omega\, \omega \coth\!\left(\frac{\hbar\omega}{2k_{B}T}\right)
\Re\!\!
\sideset{}{'}\sum_{lm}
(r_{l,21}^{\rm TE}+r_{l,21}^{\rm TM})\\
\times\left[\mathbf{\hank M}_{lm}
(\mathbf{r};k_{2})\!\otimes\!
\mathbf{\hank M}_{lm}^{\ccangles}(\mathbf{r};k_{2})\!+\!
\mathbf{\hank N}_{lm}(\mathbf{r};k_{2})\!\otimes\!
\mathbf{\hank N}_{lm}^{\ccangles}(\mathbf{r};k_{2})
\right].
\end{multline}
By symmetry, the relevant stress
tensor element for the spherical setup is
\begin{equation}
\label{4.2}
\uv{r}\tensor{T}\uv{r}=\uv{r}\tensor{\Theta}\uv{r}-
{\rm Tr}\tensor{\Theta}/2
\end{equation}
($\uv{r}$ $\!=$ $\!\mathbf{r}/r$).
By taking into account that the addition theorem (\ref{3.20})
implies the relation
\begin{equation}
\label{4.2a}
\sum_{m=-l}^{l}
\left|Y_{lm}(\theta,\phi)\right|^2
=\frac{2l+1}{4\pi}\,,
\end{equation}
and making use of Eqs.~(\ref{3.4}) and (\ref{3.5})
it is not difficult to see that the relations
\begin{equation}
\label{4.3}
\uv{r}\mathbf{\hank M}_{lm}
(\mathbf{r};k)\otimes
\mathbf{\hank M}_{lm}^{\ccangles}(\mathbf{r};k)\uv{r}=0,
\end{equation}
\begin{eqnarray}
\label{4.4}
\lefteqn{
\sum_{m}\uv{r}\mathbf{\hank N}_{lm}(\mathbf{r};k)\otimes
\mathbf{\hank
N}_{lm}^{\ccangles}(\mathbf{r};k)\uv{r}
}
\nonumber\\&&\hspace{5ex}
=(2l+1)\frac{\hank
h_{l}^{2}(kr)}{2\pi^2}\, \frac{l(l+1)}{r^2}\,,
\end{eqnarray}
\begin{equation}
\label{4.5}
\sum_{m}
\mathbf{\hank M}_{lm}
(\mathbf{r};k)
\mathbf{\hank M}_{lm}^{\ccangles}(\mathbf{r};k)=(2l+1)\frac{\hank
h_{l}^{2}(kr)}{2\pi^2}\,k^2,
\end{equation}
and
\begin{eqnarray}
\label{4.6}
\lefteqn{
\sum_{m}
\mathbf{\hank N}_{lm}
(\mathbf{r};k)
\mathbf{\hank N}_{lm}^{\ccangles}(\mathbf{r};k)
}
\nonumber\\&&
\!\!\!
=\frac{2l+1}{2\pi^2 r^2}
\left\{l(l+1)\,\hank
h_{l}^{2}(kr)+\left[kr \hank h'_{l}(kr)+
\hank h_{l}(kr)\right]^2\right\}
\qquad
\end{eqnarray}
are valid. Substituting Eqs.~(\ref{4.3})--(\ref{4.6}) in Eq.~(\ref{4.1}), we
obtain from Eq.~(\ref{4.2}) the Casimir force per unit area
on the surface of a sphere of radius $R$:
\begin{eqnarray}
\label{4.7}
\lefteqn{
T_{RR}=\frac{\hbar}{8\pi^2 c} \int_{0}^\infty
d\omega\, \omega \coth\!\left(\frac{\hbar\omega}{2k_{B}T}\right)
}
\nonumber\\&&\hspace{5ex}
\times \,
\Re\sum_{l=1}^{\infty}(2l+1)
\left(r_{l,21}^{\rm TE}+r_{l,21}^{\rm TM}\right)F_{l},
\end{eqnarray}
where
\begin{eqnarray}
\label{4.8}
\lefteqn{
F_{l}=
\frac{l(l+1)}{R^2}\,\hank h_{l}^{2}(k_{2}R)
-k_{2}^2\hank h_{l}^{2}(k_{2}R)
}
\nonumber\\&&\hspace{5ex}
-\frac{1}{R^2} \left[k_{2}R \hank h'_{l}(k_{2}R)+
\hank h_{l}(k_{2}R)
\right]^2.
\end{eqnarray}
Note that the frequency integral can
be further processed by the
standard contour deformation to the imaginary frequency
axis, thereby leading to a sum
over residues (Matsubara frequencies) when \mbox{$T$ $\!>$ $\!0$}. 
\section{Summary}
We have derived a general
expression [Eqs.~(\ref{2.11}) and (\ref{2.12})]
for the Casimir stress that is valid
for arbitrary dispersing and absorbing magnetodielectric material,
and have evaluated it for (layered) spheres. In particular,
Eq.~(\ref{4.7}) [together with Eq.~(\ref{4.8})] is applicable to
spheres made of left-handed material, where interesting
results can be expected when the relevant Mie resonances
correspond to spectral regions
in which the material behaves left-handed. Since the 
outlined Green tensor construction also allows for linear amplification,
the theory of Casimir forces may also be extended to
(linearly) amplifying bodies. For this purpose, the calculations
can again be based on the quantization scheme
given in Ref.~\cite{Welsch},
if appropriate modifications (of the definitions
of the fundamental bosonic fields) are taken into account.
Such modifications will effectively add to Eq.~(\ref{2.12})
another integral, over
those frequency intervals and spatial regions which
can contribute to the amplification.
%
%%%%%%%%%%%%%%%%%%%%%%%%%%%%%%%%%%%%%%%%%%%%%%%%%%%%

\acknowledgments
C.R. wishes to thank
Ho Trung Dung, Stefan Scheel and Mikayel Khanbekyan
for discussions.
%%%%%%%%%%%%%%%%%%%%%%%%%%%%%%%%%%%%%%%%%%%%%%%%%%%
%
\bibliographystyle{apsrev.bst}
\bibliography{DissBib}

\begin{thebibliography}{21}
\expandafter\ifx\csname natexlab\endcsname\relax\def\natexlab#1{#1}\fi
\expandafter\ifx\csname bibnamefont\endcsname\relax
  \def\bibnamefont#1{#1}\fi
\expandafter\ifx\csname bibfnamefont\endcsname\relax
  \def\bibfnamefont#1{#1}\fi
\expandafter\ifx\csname citenamefont\endcsname\relax
  \def\citenamefont#1{#1}\fi
\expandafter\ifx\csname url\endcsname\relax
  \def\url#1{\texttt{#1}}\fi
\expandafter\ifx\csname urlprefix\endcsname\relax\def\urlprefix{URL }\fi
\providecommand{\bibinfo}[2]{#2}
\providecommand{\eprint}[2][]{\url{#2}}

\bibitem[{\citenamefont{Boyer}(1968)}]{Boyer}
\bibinfo{author}{\bibfnamefont{T.~H.} \bibnamefont{Boyer}},
  \bibinfo{journal}{Phys. Rev.} \textbf{\bibinfo{volume}{174}},
  \bibinfo{pages}{1764} (\bibinfo{year}{1968}).

\bibitem[{\citenamefont{Casimir}(1956)}]{CasimirElectronModel}
\bibinfo{author}{\bibfnamefont{H.~B.~G.} \bibnamefont{Casimir}},
  \bibinfo{journal}{Physica} \textbf{\bibinfo{volume}{19}},
  \bibinfo{pages}{846} (\bibinfo{year}{1956}).

\bibitem[{\citenamefont{Schwinger}(1993)}]{SchwingerSonoluminescence}
\bibinfo{author}{\bibfnamefont{J.}~\bibnamefont{Schwinger}},
  \bibinfo{journal}{Proc. Natl. Acad. Sci. USA} \textbf{\bibinfo{volume}{90}},
  \bibinfo{pages}{958,2105,4505,7285} (\bibinfo{year}{1993}).

\bibitem[{\citenamefont{Frenzel and Schultes}(1934)}]{SonoluminescenceFirst}
\bibinfo{author}{\bibfnamefont{H.}~\bibnamefont{Frenzel}} \bibnamefont{and}
  \bibinfo{author}{\bibfnamefont{H.}~\bibnamefont{Schultes}},
  \bibinfo{journal}{Z. Phys. Chem.} \textbf{\bibinfo{volume}{27}},
  \bibinfo{pages}{421} (\bibinfo{year}{1934}).

\bibitem[{\citenamefont{Barber et~al.}(1997)\citenamefont{Barber, Hiller,
  L\"ofstedt, Putterman, and Weniger}}]{SonoluminescenceReview}
\bibinfo{author}{\bibfnamefont{B.~P.} \bibnamefont{Barber}},
  \bibinfo{author}{\bibfnamefont{R.~A.} \bibnamefont{Hiller}},
  \bibinfo{author}{\bibfnamefont{R.}~\bibnamefont{L\"ofstedt}},
  \bibinfo{author}{\bibfnamefont{S.~J.} \bibnamefont{Putterman}},
  \bibnamefont{and} \bibinfo{author}{\bibfnamefont{K.}~\bibnamefont{Weniger}},
  \bibinfo{journal}{Phys. Rep.} \textbf{\bibinfo{volume}{281}},
  \bibinfo{pages}{65} (\bibinfo{year}{1997}).

\bibitem[{\citenamefont{Milton and Ng}(1997)}]{MiltonSonoluminescence}
\bibinfo{author}{\bibfnamefont{K.~A.} \bibnamefont{Milton}} \bibnamefont{and}
  \bibinfo{author}{\bibfnamefont{Y.~J.} \bibnamefont{Ng}},
  \bibinfo{journal}{Phys. Rev. E} \textbf{\bibinfo{volume}{55}},
  \bibinfo{pages}{4207} (\bibinfo{year}{1997}).

\bibitem[{\citenamefont{Barton}(1999)}]{Barton}
\bibinfo{author}{\bibfnamefont{G.}~\bibnamefont{Barton}}, \bibinfo{journal}{J.
  Phys. A Math. Gen.} \textbf{\bibinfo{volume}{32}}, \bibinfo{pages}{525}
  (\bibinfo{year}{1999}).

\bibitem[{\citenamefont{Klich}(2000)}]{Klich1}
\bibinfo{author}{\bibfnamefont{I.}~\bibnamefont{Klich}},
  \bibinfo{journal}{Phys. Rev. D} \textbf{\bibinfo{volume}{61}},
  \bibinfo{pages}{025004} (\bibinfo{year}{2000}).

\bibitem[{\citenamefont{Klich et~al.}(2000)\citenamefont{Klich, Feinberg, Mann,
  and Revzen}}]{Klich2}
\bibinfo{author}{\bibfnamefont{I.}~\bibnamefont{Klich}},
  \bibinfo{author}{\bibfnamefont{J.}~\bibnamefont{Feinberg}},
  \bibinfo{author}{\bibfnamefont{A.}~\bibnamefont{Mann}}, \bibnamefont{and}
  \bibinfo{author}{\bibfnamefont{M.}~\bibnamefont{Revzen}},
  \bibinfo{journal}{Phys. Rev. D} \textbf{\bibinfo{volume}{62}},
  \bibinfo{pages}{045017} (\bibinfo{year}{2000}).

\bibitem[{\citenamefont{Brevik and Yousef}(2000)}]{Brevik}
\bibinfo{author}{\bibfnamefont{I.}~\bibnamefont{Brevik}} \bibnamefont{and}
  \bibinfo{author}{\bibfnamefont{T.~A.} \bibnamefont{Yousef}},
  \bibinfo{journal}{J. Phys. A Math. Gen.} \textbf{\bibinfo{volume}{33}},
  \bibinfo{pages}{5819} (\bibinfo{year}{2000}).

\bibitem[{\citenamefont{Nesterenko}(2001)}]{Nesterenko2}
\bibinfo{author}{\bibfnamefont{V.~V.} \bibnamefont{Nesterenko}},
  \bibinfo{journal}{Phys. Rev. D} \textbf{\bibinfo{volume}{64}},
  \bibinfo{pages}{025013} (\bibinfo{year}{2001}).

\bibitem[{\citenamefont{Milton}(1997)}]{MiltonDdim}
\bibinfo{author}{\bibfnamefont{K.~A.} \bibnamefont{Milton}},
  \bibinfo{journal}{Phys. Rev. D} \textbf{\bibinfo{volume}{55}},
  \bibinfo{pages}{4940} (\bibinfo{year}{1997}).

\bibitem[{\citenamefont{Nesterenko and Pirozhenko}(1998)}]{Nesterenko}
\bibinfo{author}{\bibfnamefont{V.~V.} \bibnamefont{Nesterenko}}
  \bibnamefont{and} \bibinfo{author}{\bibfnamefont{I.~G.}
  \bibnamefont{Pirozhenko}}, \bibinfo{journal}{Phys. Rev. D}
  \textbf{\bibinfo{volume}{57}}, \bibinfo{pages}{1284} (\bibinfo{year}{1998}).

\bibitem[{\citenamefont{Kn\"{o}ll et~al.}(2001)\citenamefont{Kn\"{o}ll, Scheel,
  and Welsch}}]{Welsch}
\bibinfo{author}{\bibfnamefont{L.}~\bibnamefont{Kn\"{o}ll}},
  \bibinfo{author}{\bibfnamefont{S.}~\bibnamefont{Scheel}}, \bibnamefont{and}
  \bibinfo{author}{\bibfnamefont{D.-G.} \bibnamefont{Welsch}},
  \emph{\bibinfo{title}{Coherence and Statistics of Photons and Atoms}}
  (\bibinfo{publisher}{Wiley}, \bibinfo{address}{New York},
  \bibinfo{year}{2001}), chap.~\bibinfo{chapter}{1}, Wiley Series in Lasers and
  Applications, \bibinfo{note}{note recent erratum quant-ph/0003121}.

\bibitem[{\citenamefont{Raabe et~al.}(2003)\citenamefont{Raabe, Kn\"{o}ll, and
  Welsch}}]{Raabe1}
\bibinfo{author}{\bibfnamefont{C.}~\bibnamefont{Raabe}},
  \bibinfo{author}{\bibfnamefont{L.}~\bibnamefont{Kn\"{o}ll}},
  \bibnamefont{and} \bibinfo{author}{\bibfnamefont{D.-G.}
  \bibnamefont{Welsch}}, \bibinfo{journal}{Phys. Rev. A}
  (\bibinfo{year}{2003}), \bibinfo{note}{in production, also quant-ph/0212154}.

\bibitem[{\citenamefont{Mie}(1908)}]{Mie}
\bibinfo{author}{\bibfnamefont{G.}~\bibnamefont{Mie}}, \bibinfo{journal}{Ann.
  d. Physik} \textbf{\bibinfo{volume}{25}}, \bibinfo{pages}{377}
  (\bibinfo{year}{1908}).

\bibitem[{\citenamefont{Chew}(1995)}]{ChewBook}
\bibinfo{author}{\bibfnamefont{W.~C.} \bibnamefont{Chew}},
  \emph{\bibinfo{title}{Waves and Fields in inhomogeneous Media}}, IEEE Press
  Series on Electromagnetic Waves (\bibinfo{publisher}{IEEE Press},
  \bibinfo{address}{New York}, \bibinfo{year}{1995}).

\bibitem[{\citenamefont{Born and Wolf}(1998)}]{BornWolf}
\bibinfo{author}{\bibfnamefont{M.}~\bibnamefont{Born}} \bibnamefont{and}
  \bibinfo{author}{\bibfnamefont{E.}~\bibnamefont{Wolf}},
  \emph{\bibinfo{title}{Priciples of Optics}} (\bibinfo{publisher}{Cambridge
  University Press}, \bibinfo{address}{Cambridge, United Kingdom},
  \bibinfo{year}{1998}), \bibinfo{edition}{sixth (corrected)} ed.

\bibitem[{\citenamefont{Jackson}(1983)}]{Jackson}
\bibinfo{author}{\bibfnamefont{J.~D.} \bibnamefont{Jackson}},
  \emph{\bibinfo{title}{Klassische Elektrodynamik}} (\bibinfo{publisher}{Walter
  de Gruyter}, \bibinfo{address}{Berlin, New York}, \bibinfo{year}{1983}),
  \bibinfo{edition}{second, improved} ed.

\bibitem[{\citenamefont{Tai}(1993)}]{Taibook}
\bibinfo{author}{\bibfnamefont{C.-T.} \bibnamefont{Tai}},
  \emph{\bibinfo{title}{Dyadic Green functions in electromagnetic theory}},
  IEEE Press Series on Electromagnetic Waves (\bibinfo{publisher}{IEEE Press},
  \bibinfo{address}{New York}, \bibinfo{year}{1993}), \bibinfo{edition}{2nd}
  ed.

\bibitem[{\citenamefont{Morse and Feshbach}(1953)}]{MorseAndFeshbach}
\bibinfo{author}{\bibfnamefont{P.~M.} \bibnamefont{Morse}} \bibnamefont{and}
  \bibinfo{author}{\bibfnamefont{H.}~\bibnamefont{Feshbach}},
  \emph{\bibinfo{title}{Methods of theoretical physics}}, vol.
  \bibinfo{volume}{1 and 2} of \emph{\bibinfo{series}{International Series In
  Pure And Applied Physics}} (\bibinfo{publisher}{McGraw-Hill Book Company
  Inc.}, \bibinfo{address}{New York}, \bibinfo{year}{1953}).

\end{thebibliography}
\end{document}